\documentclass{article}
\usepackage{bm}
\usepackage{bbm}
\usepackage{xcolor}
\usepackage{pifont}
\usepackage{booktabs}
\usepackage{amsmath, amssymb, bigstrut}
\usepackage{amsthm}
\usepackage{mathtools}
\usepackage{apacite}
\usepackage{natbib}
\usepackage{graphicx}
\usepackage{adjustbox}
\usepackage{caption}
\usepackage{lscape}
\usepackage{subcaption}
\usepackage{array, makecell}
\usepackage[linesnumbered,ruled,vlined]{algorithm2e}
\usepackage[utf8]{inputenc}
\usepackage[a4paper, total={6.5in, 9in}]{geometry}

\newsavebox{\mybox}\newsavebox{\mysim}
\newcommand{\distras}[1]{%
  \savebox{\mybox}{\hbox{\kern3pt$\scriptstyle#1$\kern3pt}}%
  \savebox{\mysim}{\hbox{$\sim$}}%
  \mathbin{\overset{#1}{\kern\z@\resizebox{\wd\mybox}{\ht\mysim}{$\sim$}}}%
}
\makeatother

\usepackage[colorlinks=true,citecolor=blue,allcolors=blue]{hyperref} 

\title{\textbf{To Vary or Not To Vary: A Flexible Empirical Bayes Factor for Testing Variance Components}}
\author{{{Fabio Vieira}$^{1}$\footnote{Email: f.gen.vieira@gmail.com}} \and Hongwei Zhao$^2$ \and Joris Mulder$^1$ }
\date{%
     $^1$Department of Methodology and Statistics, Tilburg University\\%
     $^2$Unit of Quantitative Psychology and Individual Differences, KU Leuven\\[2ex]%
     \today
}

\setcitestyle{aysep={}} 
\providecommand{\keywords}[1]
{
  \small	
  \textbf{Keywords:} #1
}

\newcolumntype{R}{@{\extracolsep{0.2cm}}r@{\extracolsep{10pt}}}%

\begin{document}

\RestyleAlgo{ruled}

\maketitle

\begin{abstract}
Random effects are the gold standard for capturing structural heterogeneity in data, such as spatial dependencies, individual differences, or temporal dependencies. However, testing for their presence is challenging, as it involves a variance component constrained to be non-negative -- a boundary problem. This paper proposes a flexible empirical Bayes factor (EBF) for testing random effects. Rather than testing whether a variance component is zero, the EBF tests the equivalent hypothesis that all random effects are zero. Crucially, it avoids manual prior specification based on external knowledge, as the distribution of random effects is part of the model’s lower level and estimated from the data—yielding an “empirical” Bayes factor. The EBF uses a Savage-Dickey density ratio, allowing all random effects to be tested using only the full model fit. This eliminates the need to fit multiple models with different combinations of random effects. Simulations on synthetic data evaluate the criterion’s general behavior. To demonstrate its flexibility, the EBF is applied to generalized linear crossed mixed models, spatial random effects models, dynamic structural equation models, random intercept cross-lagged panel models, and nonlinear mixed effects models.
\end{abstract}


\section{Introduction}

\par
Structural heterogeneity in the data can be caused by different potential sources, such as temporal dependencies, individual differences, or spatial dependencies. Random effects models are the gold standard for capturing the heterogeneity caused by these different sources where the corresponding variance components quantify the degree of the heterogeneity. Random effects models have resulted in numerous insights about the degree of statistical heterogeneity in empirical data. Examples include clinical trials \citep{cleophas2008random}, economics \citep{menegaki2011growth}, finance \citep{alsakka2010random}, social networks \citep{vieira2024fast}, meta-analysis \citep{borenstein2010basic}, among others. 

\par
A crucial problem when building random-effect models is to test whether the heterogeneity that is caused by the different sources is sufficient for random effects to be incorporated into the model.
This testing problem serves two purposes. First, understanding the degree of heterogeneity caused by the different sources is important from a substantive perspective. As an example, \cite{hamaker2011model} identified individual differences regarding psychological constructs in longitudinal research using random effects. Second, testing the presence of random effects is of crucial importance when building statistical models for data with complex hierarchical or dependency structures. If a potential source of heterogeneity actually has a negligible impact on the heterogeneity in the data, it is recommended to omit the random effect to avoid statistical overfitting. 

\par
Over the years, many different methods have been proposed to carry out this test \citep[e.g.,][among many others]{crainiceanu2008likelihood, zhang2008variance, dunson2008random, verhagen2013bayesian, roback2021beyond, pauler1999bayes, du2020testing}. A few examples of existing methods include marginal likelihood approaches (including Bayes factors), information criteria, likelihood ratio tests, $k$-fold cross-validation methods, or simply eyeballing the estimates of the variance components. These methods have different potential limitations, such as subjectivity (e.g., eyeballing), computational complexity (e.g., computing marginal likelihoods, cross-validation), problems to control error rates (significance tests), or complex tuning (e.g., choice of the prior in marginal likelihood approaches).

\par
This paper proposes an empirical Bayes factor (EBF) for testing random effects as an alternative to the above approaches with their inherent potential limitations. To circumvent the heavy dependence of the Bayes factor on the prior based on external/contextual information, which may be particularly challenging for random effects variances, an empirical Bayesian approach is considered \citep{casella1985introduction,ten1999empirical,liu2019use}. We achieve this using the fact that the distribution of the random effects on the lower level effectively serves as a prior for the random effects, which is estimated from the data when fitting the model. Moreover, to avoid the need to compute marginal likelihoods, which is the main building block of Bayes factors, a Savage-Dickey density ratio is considered \citep{dickey1971weighted}. Due to the Savage-Dickey density approach, it is also possible to test all separate random effects by only requiring the full fitted model. Alternative approaches, such as information criteria and cross-validation methods, on the other hand, require fitting all possible models with all possible combinations of random effects, which can be computationally very demanding, limiting their applicability to larger, complex random effects models (such as dynamic structural equation modeling). On the other hand, these alternatives methods are not restricted to testing nested models, which is the case for the proposed EBF which is specifically designed as a diagnostic check of whether a specific random effect (or multiple random effects) should be included or excluded. Importantly, the EBF is designed for random effects with any covariance structure, thereby extending the method of \cite{vieira2023bayesian}.


\par
A unique aspect of the proposed test is that it only requires an estimate of the variance components under the full model and an estimate and error covariance matrix of the random effects of interest. These elements can simply be obtained using MCMC output after fitting the full model. Subsequently, the EBF can be computed instantly without requiring additional computer intensive sampling procedures. This makes the test very flexible and fast for testing random effects once a model has been fitted. 
An R package \textbf{EBF} was developed with the method proposed and it is available on the link \url{https://github.com/Fabio-Vieira/EBF}. Implementation in \textbf{BFpack} \citep{mulder2021bfpack} is planned for the near future.

 

\par
The remainder of this paper is structured as follows. Section \ref{sec:alternatives} gives a brief description of existing alternatives to conduct this test. Section \ref{sec:EBF} introduces the EBF. Section \ref{sec:simulation} presents a synthetic data study showing the behavior of the EBF. Subsequently, Section \ref{sec:application} illustrates the applicability of the EBF to different modeling scenarios. Finally, Section \ref{sec:discussion} ends up with concluding remarks and a discussion.

\section{Existing methods for testing random effects}\label{sec:alternatives}
This section discusses common existing ways for the testing problem of testing variance components in random effects models including their potential shortcomings. Table \ref{tab:method_comp} provides an overview of methods which will be briefly discussed below. Note that this list is not intended to be an exhaustive overview of every possible method that has been proposed in the literature but mainly focusses on the most common methods.

\begin{table}[t]
    \centering
    \begin{tabular}{lllll}
    \hline
    Method & Aim & Computation & Tuning\\
    \hline
    Eyeballing  &  Subjectively evaluate $\hat{\tau}^2 $ & None & None\\
    Information criteria & Assess model adequacy by & Low to moderate & \#parameters,\\
    & balancing fit and complexity & & \#observations \\
    Likelihood ratio tests & Check significance (bootstrapping) & High & Sign. level $\alpha$ \\
    $K$-fold cross-validation & Assess predictive performance & Very high & Choice of $K$ \\
    Marginal modeling  & Evaluate $P(\tau^2 > 0 | \text{Data})$ & Low for simple designs & None \\
    Bayes factors & Assess model adequacy & Very high & Prior for $\tau^2$ \\
    Empirical Bayes factor & Test $\tau^2 = 0$ vs $\tau^2 > 0$ & Low (only fit full model) & None\\
    \hline
    \end{tabular}
    \caption{Overview of methods for testing a random effects variance $\tau^2$.}
    \label{tab:method_comp}
\end{table}

One practical approach is eyeballing \citep{tukey1977exploratory, hartwig1979exploratory}. When assessing variance components, eyeballing entails drawing inferences by visually inspecting parameter estimates or confidence intervals of the variance components without a formal statistical framework \citep{ludbrook2011there}. If the (interval) estimate is deemed close enough to zero, then it would be concluded that the corresponding random effect should be excluded from the model. While this may serve as a quick and easy exploratory tool, eyeballing is also prone to the subjective interpretation of the researcher. Because variance parameters in multilevel models are inherently positive, the method cannot effectively test whether heterogeneity is absent, potentially resulting in misleading or pseudo-scientific conclusions \citep{sharpe2015chi}.

The broad class of information criteria is another popular methodology for this purpose. Examples include the Akaike information criterion \citep{akaike1998information}, the Bayesian information criterion \citep{schwarz1978estimating,raftery1995bayesian}, and the deviance information criterion \citep{spiegelhalter2002bayesian}. These criteria balance model fit and complexity, but their application in random effects models may not be straightforward. For example, it is unclear whether the random effects count as free parameters or not when computing model complexity. For the BIC, the sample size is required but it is unclear whether all observations should be equally counted given the potential dependencies in the data \citep{berger2014effective}. Moreover, obtaining reliable maximum likelihood estimates in complex or high-dimensional models is difficult. Methods to estimate the effective number of parameter in the DIC can be numerically unstable, undermining their applicability \citep{gelman2014understanding}. For additional references on information criteria see \cite{celeux2006deviance, millar2018conditional, merkle2019bayesian, zhang2019bayesian}. 

Likelihood ratio tests (LRTs) compare nested models based on differences in log-likelihoods, but the standard asymptotic assumptions (typically resulting in a mixture of chi-squared distributions) often fail when testing variance components on the boundary of their parametric space \citep{stram1994variance, leamer2010tantalus, abbott2022far, crainiceanu2008likelihood, zhang2008variance}. This failure leads to unreliable type I error control, and practitioners usually resort to computationally expensive parametric bootstrapping to approximate the correct distribution — rendering the approach less practical for routine analysis \citep{roback2021beyond}.

$K$-fold cross-validation methods assess the out-of-sample predictive performance by partitioning the data into training and test sets \citep{berrar2019cross}. This methods requires an arbitrary choice of $K$. An increasingly popular choice is to set $K$ equal to the number of observations leading to what is called leave-one-out cross-validation. Though powerful when evaluating predictive performance, there has been some debate about its inconsistent behavior, failing to endorse the simpler model even when data perfectly aligns with it \citep{shao1993linear, gronau2019limitations}. Additionally, its computational demands increase with repeated trials complicating its application for challenging random effects models.

Another approach is marginal modeling where the random effects are integrated and the variance components become covariance parameters in structured covariance matrices. As covariances can be negative, this approach circumvents the boundary problem of whether the variance is zero or positive. Though statistically elegant, to our knowledge this method has mainly been used for simpler model setups with only random intercepts \citep{mulder2013bayesian, fox2017bayes, mulder2019bayes, nielsen2021small}. In more complex models, ensuring that the integrated covariance matrix is positive definite and handling the potentially complex restrictions in a MCMC framework may pose computational challenges.

Bayes factors \citep{jeffreys1961theory} offer a fully Bayesian framework for model comparison by assessing the ratio of marginal likelihoods \citep{sinharay2002sensitivity, liu2008bayes, vanpaemel2010prior,hoijtink2019tutorial}. Yet Bayes factors face two key challenges. First, as the Bayes factor is particularly sensitive to the prior of the parameter that is tested, which is the variance component in the current paper, specifying a substantively meaningful prior may be challenging and potentially subjective. Setting an arbitrarily vague prior is also not allowed as this would result in Bartlett's paradox \citep{liang2008mixtures,jeffreys1961theory,Bartlett:1957}. Moreover, the objective Bayes factor approach of \cite{garcia2007objective} for testing a random intercept has not been extended to general random effects testing. A second challenge is the computation of the marginal likelihoods which is computationally intensive, making the approach less attractive for practical applications in case of large complex models \citep{gelman1998simulating, geweke1999using, friel2008marginal}.


Finally, a general issue of all these methods is the need to fit all possible models with all combinations of included/excluded random effects. For testing a single random effect, two models must be fitted. For $P$ potential random effects, the methods would require fitting $2^P$ models. Clearly, this can become a severe computational burden. Next we discuss an empirical Bayes factor for testing random effects which avoids this general issue, and also various inherent limitations of the above mentioned methods when testing variance components.

\section{The empirical Bayes factor}\label{sec:EBF}
We write the model selection problem of whether a vector of $J$ random effects, say, $\bm\theta$, is included or excluded as follows:
\begin{eqnarray}
    \label{eq:model}
    \mathcal{M}_1&:&
    \left\{\begin{array}{lcl}
        \bm{y} &\sim& p(\bm{X}, \bm{\phi}, \bm{\theta})\\
        \bm{\theta} &\sim& \mathcal{N}(\bm{0}, \bm{\Psi}(\bm{\tau}))\\
        \end{array}\right.\\
\nonumber    \mathcal{M}_0&:& ~~~~\bm{y}~~\, \sim ~~\, p(\bm{X}, \bm{\phi}, \textbf{0}),
\end{eqnarray}
where $\bm{\Psi}(\bm{\tau})$ is a structured covariance matrix with unknown variance components $\bm\tau$, $\bm{\phi}$ are the nuisance parameters, and $\bm{X}$ is an array with observed predictor variables. The simplest example of this model selection problem would be between a random intercept model which assumes the observations are clustered according to observed groups and a null model which assumes there is no clustering across groups. In Section \ref{sec:application}, various examples are given of different modeling scenarios with multiple random effects and latent variables.


We start by reformulating the model selection problem by omitting the nuisance elements, such that we can write the models in the following compact manner:
\begin{eqnarray}
    \label{eq:model1}
    \mathcal{M}_1 &:& \bm{\theta}\in\mathbb{R}^J\\
\nonumber    \mathcal{M}_0 &:& \bm{\theta}=\textbf{0}.
\end{eqnarray}
When assuming that the distribution of the random effects in the lower level of $\mathcal{M}_1$ would be known, this distribution could be viewed as prior distribution of $\bm{\theta}$. Consequently, when assessing whether the vector $\bm{\theta}$ is equal to a constant vector as in \eqref{eq:model1}, the Bayes factor of the null model $\mathcal{M}_0$ against the full model $\mathcal{M}_1$ can be expressed as a Savage-Dickey density ratio \citep{dickey1971weighted}:
\begin{eqnarray}
\label{SDBF}
BF_{01} = \frac{p({\bm\theta}= \bm 0 | \bm{y},\mathcal{M}_1)}{p({\bm\theta}=\bm 0|\mathcal{M}_1)},
\end{eqnarray}
which should be read as the density of the posterior of $\bm\theta$ at the constant $\bm 0$ under the full model $\mathcal{M}_1$ divided by the density of the prior of $\bm\theta$ at the null vector $\bm 0$ under the full model $\mathcal{M}_1$. This expression of the Bayes factor also clearly shows its sensitivity to the prior of the parameters that we tested: When specifying the prior for $\bm\theta$ under $\mathcal{M}_1$ arbitrarily vague, the Bayes factor $B_{01}$ becomes arbitrarily large. 

Importantly, this `prior distribution' of the random effects $\bm\theta$ is not a prior in a strict Bayesian sense but it is actually the lower level part of the mixed effects model, which is fitted using the observed data. Therefore, we can use the available empirical data to estimate the random effects distribution in the denominator in \eqref{SDBF}. To achieve this, we can use the full posterior for the variance components, i.e.,
\begin{eqnarray}
\nonumber p({\bm\theta}=\bm 0|\mathcal{M}_1) &=& \int
\mathcal{N}(\bm{0}|\bm{0}, \bm{\Psi}(\bm\tau))p(\bm\tau|\textbf{y})d\bm\tau\\
\nonumber &\approx& S^{-1}\sum_{s=1}^S \mathcal{N}(\bm{0}|\bm{0}, \bm{\Psi}(\bm\tau^{(s)}))\\
\label{fullpost} &=& (2\pi)^{-J/2}S^{-1}\sum_{s=1}^S |\bm{\Psi}(\bm\tau^{(s)})|^{-1/2},
\label{estprior}
\end{eqnarray}
where $\bm\tau^{(s)}$ denotes the $s$-th draw of the marginal posterior of the variance components $\bm\tau$.
Instead of using the full posterior, empirical Bayesian approaches are often based on a single point estimate \citep{casella1985introduction,george2000calibration,mulder2021prevalence}, where the posterior uncertainty is ignored. Thus, we could also approximate the denominator in \eqref{SDBF} by
\begin{equation}
p({\bm\theta}=\bm 0|\mathcal{M}_1) \approx
\mathcal{N}(\bm{0}|\bm{0}, \bm{\Psi}(\bar{\bm\tau}))=(2\pi)^{-J/2}|\bm{\Psi}(\bar{\bm\tau})|^{-1/2},
\label{estprior2}
\end{equation}
where $\bar{\bm\tau}$ denotes a point estimate for the variance components. A natural choice for the point estimate in empirical Bayesian approaches is to use the posterior mode. For the current problem however, the use of the posterior mode can be problematic as the posterior mode of a variance component under a mixed effects model can be zero, similar as the maximum likelihood estimate. Plugging in a zero estimate for the variance components would result in a Bayes factor of $B_{01}=0$ in \eqref{SDBF}, implying a complete bias towards the full model. For the same reason, the use of the full posterior in \eqref{fullpost} to obtain the EBF may also result in a bias towards the full model when the posterior of the random effects variance is peaked at zero (as we shall see in the simulations). For this reason, we require an estimate for the random effects variance that is strictly larger than zero. Because the posterior mean $\bar{\tau}$ satisfies this property, we shall use this plug-in estimate for the variance components to compute the EBF via \eqref{estprior2}. 

In addition to the denominator in \eqref{SDBF}, we also need to the posterior density at $\bm\theta=\textbf{0}$ under the full model $\mathcal{M}_1$in the numerator of \eqref{SDBF}. To estimate this quantity from a random posterior sample, which is typically available after fitting a Bayesian model, we first approximate the posterior of the random effects with a Gaussian distribution \citep{raftery1995bayesian,gu2018approximated,williams2020comparing}. Subsequently, we estimate the posterior density via the approximated Gaussian distribution, i.e.,
\begin{equation}
p(\bm{\theta}=\textbf{0} | \bm{y},\mathcal{M}_1) \approx \mathcal{N}(\textbf{0}|\bar{\bm{\theta}}, \bar{\bm{\Sigma}}_{\bm\theta})= (2\pi)^{-J/2} |\bar{\bm{\Sigma}}_{\bm\theta}|^{-1/2}\exp\{-\tfrac{1}{2}\bar{\bm{\theta}}'\bar{\bm{\Sigma}}^{-1}_{\bm\theta}\bar{\bm{\theta}}\},
\end{equation}
where $\bar{\bm{\theta}}$ denotes the posterior mean of the random effects and $\bar{\bm{\Sigma}}$ denotes its covariance matrix. Consequently, the empirical Bayes factor can be computed as
\begin{eqnarray}
\nonumber
\text{EBF}_{01} &=& \frac{\pi({\bm\theta}= \bm 0 | \bm{y},\mathcal{M}_1)}{\pi({\bm\theta}=\bm 0|\mathcal{M}_1)}\\
&=&
\left\{\begin{array}{ll} |\bar{\bm{\Sigma}}_{\theta}|^{-1/2}
\exp\{-\tfrac{1}{2}\bar{\bm{\theta}}'\bar{\bm{\Sigma}}_{\theta}^{-1}\bar{\bm{\theta}}\}|\bm{\Psi}(\bar{{\tau}})|^{1/2}&
\text{ using the posterior mean,}\\
|\bar{\bm{\Sigma}}_{\theta}|^{-1/2}
\exp\{-\tfrac{1}{2}\bar{\bm{\theta}}'\bar{\bm{\Sigma}}_{\theta}^{-1}\bar{\bm{\theta}}\} 
S^{-1}\sum_{s=1}^S |\bm{\Psi}(\bm\tau^{(s)})|^{-1/2}
&\text{ using the full marginal posterior.}
\end{array}
\right.
\label{EBF01final}
\end{eqnarray}

Although Bayes factors are known to be sensitive to the choice of the prior, this is not as big of an issue for the empirical Bayes factor. The reason is that we only need the posterior for the variance components, which comes down to an estimation problem where prior sensitivity plays a much smaller role. Throughout this paper, we use flat noninformative priors to avoid (some slight) prior shrinkage in a certain (arbitrary) direction which could occur when using vague proper priors. For the variance of a random intercept $\tau^2$, this improper prior can be written as $\pi(\tau^2)\propto 1$, which was recommended by \cite{chung2013variance} as a default choice in estimation problems\footnote{It can be shown that the improper gamma(2,0) prior for the random effects standard deviation $\tau$, as recommended by \cite{chung2013variance}, is equivalent to a flat prior on the random effects variance $\tau^2$.}. Because the variance components essentially serve as nuisance parameters in the model comparison problem in \eqref{eq:model1}, noninformative improper priors can be used for these parameters.

Although the empirical Bayes factor is not as sensitive to the choice of the prior as regular Bayes factors, the specification of stable priors for variance components in mixed effects models has been an active topic in the Bayesian literature \citep{spiegelhalter2001bayesian,browne2006comparison,gelman2006prior}. For instance, noninformative Jeffreys priors should not be used for lower level variance components \citep{polson2012half}, as well as proper approximations of Jeffreys priors, such as very vague inverse Wishart priors or inverse gamma priors \citep{berger2006case}. Therefore, careful assessment of MCMC converges is very important when fitting mixed effects models. In all our analyses, the use of flat improper priors for the variance components resulted in stable posterior estimates. This will be illustrated in the application sections.

Even though empirical Bayesian approaches have been proposed for Bayes factors, to our knowledge these have mainly been proposed for testing fixed effects. A well-known example is the Bayes factor based on the $g$-prior of \cite{zellner1986assessing} where the prior hyperparameter $g$ is chosen such that it maximizes the marginal likelihood \citep[e.g.,][]{liang2008mixtures}. The current approach is fundamentally different as our `prior' is not a prior in a strict sense, such as the $g$-prior, but part of the mixed effects model, and thus, the EBF could also be computed using a classical estimate for the random effects variance. We come back to this in the Discussion. 

To obtain the required ingredients to compute the empirical Bayes factor, we need a posterior sample from the fitted mixed effects model. Depending on the employed model, different software packages can be used. The R package \texttt{brms} \citep{burkner2017brms} could be used for fitting a large variety types of mixed effects models. Many more packages are available as well for fitting specific mixed effects models. Moreover, general packages such as \texttt{rstan} \citep{guo2020package} or \texttt{rjags} \citep{plummer2012jags} allow researchers to fit virtually any type of mixed effects model. Moreover, commercial software pacakges such as Mplus or Stata, now also support Bayesian estimation algorithms from which the empirical Bayes factor can be computed.

Finally, we highlight some properties of the proposed empirical Bayes factor in relation to existing approaches. First, the proposed Bayes factor also relies on a Gaussian approximation similar as the BIC. The BIC however uses a Gaussian approximation for the entire likelihood \citep{raftery1995bayesian}, while the proposed EBF only uses a Gaussian approximation of the posterior of the (unbounded) random effects. Thus, the proposed empirical Bayes factor relies less heavily on the Gaussian approximation than the BIC. Second, unlike regular Bayes factors which are highly sensitive to the choice of the prior of the tested parameters, the EBF avoids this prior sensitivity as the `prior' is estimated from the data when fitting the full model. Finally and most importantly, in contrast to the most common methods (Table \ref{tab:method_comp}) which require fitting many different models with different combinations of included/excluded random effects, EBFs can be computed for all separate random effects once the full model has been fitted. This property drastically reduces the computational burden when testing random effects in complex mixed effects models. 



\section{Synthetic data study}\label{sec:simulation}

A simulation was performed to assess the behavior of the EBF for testing the presence of random effects. A 2-way cross-classified random effects model was considered where each cluster dimension had a 2-dimensional random effects vector:
\begin{eqnarray*}
y_{ijk} &=& \alpha_1 + x_{11ijk}\theta_{11j}+x_{12ijk}\theta_{12k}+x_{21ijk}\theta_{21j}+x_{22ijk}\theta_{22k}+\epsilon_{ijk}\\
(\theta_{11j},\theta_{12j})' &\sim & N(\textbf{0},\bm\Psi_1(\tau_{11},\tau_{12},\rho_1))\\
(\theta_{21k},\theta_{22k})' &\sim & N(\textbf{0},\bm\Psi_2(\tau_{21},\tau_{22},\rho_2))\\
\epsilon_{ijk} &\sim& \mathcal{N}(0,\sigma^2),
\end{eqnarray*}
for the first cluster index $j=1,\ldots,J$, the second cluster index $k=1,\ldots,K$, and the within cluster index $i=1,\ldots,n$, and
\[
\bm\Psi_1(\tau_{11},\tau_{12},\rho_1) = 
\left[\begin{array}{cc}
\tau_{11}^2 & \tau_{11}\tau_{12}\rho_1\\
\tau_{11}\tau_{12}\rho_1 & \tau_{12}^2
\end{array}\right] \text{ and }
\bm\Psi_2(\tau_{21},\tau_{22},\rho_2) = 
\left[\begin{array}{cc}
\tau_{21}^2 & \tau_{21}\tau_{22}\rho_2\\
\tau_{21}\tau_{22}\rho_1 & \tau_{22}^2
\end{array}\right].
\]
We will test this full model against a model where one of the four random effects is excluded, e.g., $\mathcal{M}_1:\tau_{11}>0$ against $\mathcal{M}_0:\tau_{11}=0$, when testing the first random effect of the first crossed dimension, using the proposed EBFs in \eqref{EBF01final}. Note here that if this $\mathcal{M}_0$ would receive most support and the first random effect of the first crossed dimension would be excluded, this would automatically imply that the random effects covariance $\tau_{11}\tau_{12}\rho_1$ would be zero.

Data were generated by setting $\alpha_1=0$, $\rho_1=\rho_2=0.3$, $\tau_{12} = .5$, $\tau_{21} = .5$, $\tau_{22} = 0$, and we considered different values of $\tau_{11}$ on the grid (0.00, 0.03, 0.07, 0.13, 0.20, 0.29, 0.40, 0.55, 0.75, 0.80). To assess the effect of the number of clusters on the EBF, the number of clusters of the first dimension was set either to $J=10$, 30, or 100. The number of clusters of the second dimension was fixed to $K=20$. To assess the effect of the cluster sizes, the size was set either to $n=10$, 30, or 100. For each combination of population values and sample sizes, 2,000 data sets were generated, and EBFs were computed for testing for the presence or absence of all four separate random effects $\bm\theta_{11\cdot}$, $\bm\theta_{12\cdot}$, $\bm\theta_{21\cdot}$, and $\bm\theta_{22\cdot}$. Given the chosen population values, it was expected to obtain evidence in favor of the presence of random effects $\bm\theta_{12\cdot}$ and $\bm\theta_{21\cdot}$ (because $\tau_{12} = .5$ and $\tau_{21} = .5$), evidence in favor of absence of random effects $\bm\theta_{22\cdot}$ (because $\tau_{22}=0$), and evidence in favor of absence of random effects $\bm\theta_{11\cdot}$ when $\tau_{11}=0$ (or very small) and in favor of presence of $\bm\theta_{11\cdot}$ when $\tau_{11}$ is larger than 0. The predictor variables were sampled using independent standard normal distributions. Noninformative priors were used to fit the full model, i.e.,
\begin{eqnarray*}
p(\alpha_1) \propto 1, ~p(\bm\Psi_1) \propto 1,~ p(\bm\Psi_2) \propto 1,\text{and }
p(\sigma^2) \propto \sigma^{-2}.
\end{eqnarray*}
Note that the Jeffreys prior can safely be used for the first level variance, $\sigma^2$ \citep[e.g.,][]{polson2012half}. For each generated data set, a Gibbs sampler was used to fit the full model using 5,000 posterior draws. Using the coda package \citep{plummer2015package}, the effective sample sizes of the posterior draws for the random effects variance components were checked. Generally, these varied between 3,500 and 5,000 implying reasonably accurate estimates for the posterior means.

\begin{figure}[thp]
    \centering
    \includegraphics[width = 10cm]{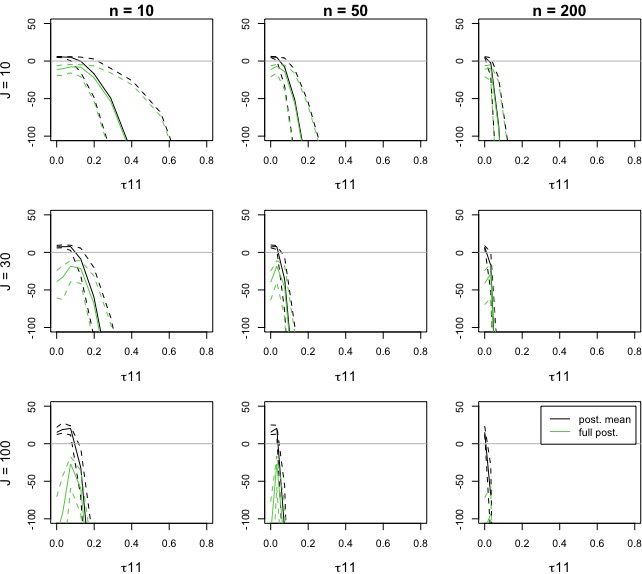}\\
    \caption{The 5\%, 50\%, and 95\% quantiles of the logarithm of the EBF in favor of absence against presence of the first random effect of the first crossed dimension (i.e., $\tau_{11}=0$ versus $\tau_{11}>0$) as a function of the true data generating $\tau_{11}$ for different cluster sizes $n$ and different numbers of clusters $J$ when using the posterior mean (black lines) or the full posterior (green lines) in \eqref{EBF01final}.}
    \label{fig_sim_tau11_ylim1}
\end{figure}

\begin{figure}[thp]
    \centering
    \includegraphics[width = 10cm]{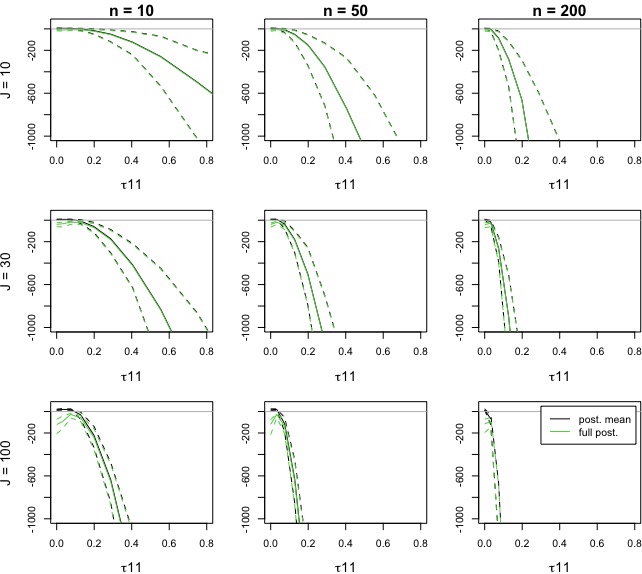}\\
    \caption{The 5\%, 50\%, and 95\% quantiles of the logarithm of the EBF in favor of absence against presence of the first random effect of the first crossed dimension (i.e., $\tau_{11}=0$ versus $\tau_{11}>0$) as a function of the true data generating $\tau_{11}$ for different cluster sizes $n$ and different numbers of clusters $J$ when using the posterior mean (black lines) or the full posterior (green lines) in \eqref{EBF01final}.}
    \label{fig_sim_tau11_ylim2}
\end{figure}

Figure \ref{fig_sim_tau11_ylim1} and \ref{fig_sim_tau11_ylim2} show the 5\%, 50\%, and 95\% quantiles of the sampling distribution of the logarithm of the $EBF_{01}$ in favor of absence against presence of the first random effect of the first crossed dimension (i.e., $\tau_{11}=0$ versus $\tau_{11}>0$) as a function of the population value of $\tau_{11}$ on two different scales on the y-axes (to clearly see the behavior near 0, and the behavior on the full scale) when using the posterior mean as point estimate (black lines) or when using the full posterior for the variance component (green lines). When using the posterior mean to compute the EBF, we observe the anticipated behavior: we obtain evidence in favor of absence of a random effect when the true $\tau_{11}$ is close to zero, and vice versa. This is not the case when using the full posterior to compute the EBF. Even for $\tau_{11}=0$, the logarithm of the EBF is generally lower than 0, implying a bias towards the full model. For larger values of $\tau_{11}$, the two EBFs are virtually identical. Moreover, the plots show that larger clusters and more clusters generally result in more evidence towards the true model.

\begin{figure}[thp]
    \centering
    \includegraphics[width = 10cm]{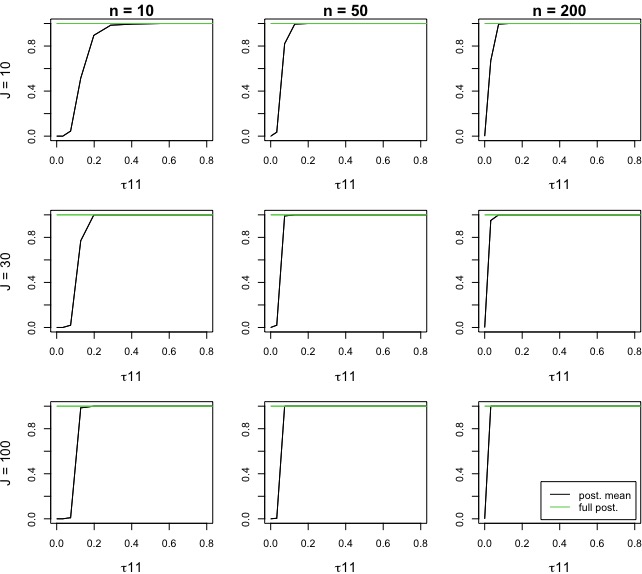}\\
    \caption{The proportion where the full model (which includes the first random effect of the first crossed dimension, corresponding to $\tau_{11}$) when using the natural cut-off value of 1 for the EBF for different values of the true data generating $\tau_{11}$ for different cluster sizes $n$ and different numbers of clusters $J$ when using the posterior mean (black lines) or the full posterior (green lines) in \eqref{EBF01final}.}
    \label{fig_sim_tau11_prop}
\end{figure}

Although the evidence as quantified by Bayes factors lies on a continuous scale, we also assess the selection behavior if we would use the EBF for dichotomous model selection. The full model is selected when $EBF_{01}<1$, and thus when $\log(EBF_{01})<0$, and the null model is selected otherwise. Figure \ref{fig_sim_tau11_prop} shows the proportion where the full model would be selected. The plots show general consistent selection behavior when the EBF is computed using the posterior mean, and a clear bias towards the full model when using the full posterior to compute the EBF. Based on these results, the use of the posterior mean to compute the EBF is generally recommended and will be used in the remainder of this paper. We come back to this in the Discussion.

When testing for the presence/absence of the other three random effects, the EBF shows similar behavior. The plots are given in Figures \ref{fig_sim_tau12}, \ref{fig_sim_tau21}, and \ref{fig_sim_tau22} in Appendix \ref{AppFig} to keep the main text as concise as possible. Finally, Figure \ref{fig_sim_tau_postmean} (Appendix \ref{AppFig}) plots the quantiles of the sampling distribution of the posterior mean of $\tau_{11}$ to see that the posterior mean is strictly larger than zero.

\section{Empirical applications}\label{sec:application}
To illustrate the flexibility of the proposed EBF and the easiness of its computation, the current section presents five model scenarios when testing random effects. In particular, we consider crossed random effects in generalized linear mixed effects models \citep{verbeke2012linear}, spatial random effects \citep{spiegelhalter2002bayesian}, crossed random effects in dynamic structural equation models \citep{asparouhov2018dynamic, mcneish2021measurement}, random intercepts in cross-lagged panel models \citep{hamaker2015critique}, and nonlinear effects mixed effects models \citep{pinheiro2000fitting}. In each application, the results are compared with the results of existing approaches which were reported in the literature. All analyses in R are available in the supplementary material of this article. Appendix \ref{app:estimates} contains the necessary posterior output when fitting the models.

\subsection{Testing crossed random effects in GLMMs}

\par
Generalized linear mixed-effect models (GLMMs) are ubiquitous in applied statistical practice to model outcome variables having various types of measurement levels \citep{mccullagh2019generalized, gelman2006data}. They allow researchers to perform analysis of data nested in groups, accounting for the variability among those groups. In particular, crossed random effects are used when different potential sources of heterogeneity are present in different dimensions in a nonnested manner \citep{verbeke2012linear}. Thus, these models provide researchers with flexible and effective framework to account for the potential variability of different grouping factors. In this illustration, we conduct an analysis to show how the empirical Bayes factor can be used to test crossed random effects in a mixed effects logistic regression model. The R code for all five applications is available in the supplementary material. Moreover, the   

\par
\cite{roback2021beyond} discuss an application, previously presented by \cite{anderson2009officiating}, where they sought evidence that officials from NCAA men's college basketball tend to even out fouls during the game. The response variable, ${y}_{ijhg}$ indicates whether a foul was called favoring the home team, which is modeled using a logistic regression model (1 if it favored the home team and 0 otherwise), where $i$ is the call, $j$ is the game, $h$ is the home team, and $g$ is the visitor team indicators. In total, there were 340 games and 4972 foul calls in the data set. It was hypothesized that this variable can be explained as a function of the foul differential which was defined by the difference between the number of foul calls received by the home and visitor teams. \cite{roback2021beyond} specified the following logistic random effects model,
\begin{equation}
\begin{gathered}
\label{eq:roback}
    y_{ijhg} \sim \text{Bernoulli}\Big(\frac{1}{1 + e^{-\mu_{ijhg}(\bm{\theta}_{1j}, \bm{\theta}_{2h}, \bm{\theta}_{3g})}}\Big)\\
    \mu_{ijhg}(\bm\theta_{1j},\bm\theta_{2h},\bm\theta_{3g})) =
    \alpha_0 + \theta_{1j1} + \theta_{2h1} + \theta_{3g1} + (\beta_0 + \theta_{1j2} + \theta_{2h2}
    + \theta_{3g2})x_{ijhg}\\
    \bm{\theta}_{1j} \sim \mathcal{N}(\bm{0}, \bm{\Psi}_1)\\
    \bm{\theta}_{2h} \sim \mathcal{N}(\bm{0}, \bm{\Psi}_2)\\
    \bm{\theta}_{3g} \sim \mathcal{N}(\bm{0}, \bm{\Psi}_3)
\end{gathered}
\end{equation}
where $\bm{\Psi}_1 = \text{diag}(\tau^2_{1},\tau^2_{1^*})$, $\bm{\Psi}_2 = \text{diag}(\tau^2_{2},\tau^2_{2^*})$ and $\bm{\Psi}_3 = \text{diag}(\tau^2_{3},\tau^2_{3^*})$. Furthermore, $x_{ijhg}$ is the foul differential for game $j = 1, \dots, 340$, home team $h = 1, \dots, 39$, visitor $g = 1, \dots, 39$, and call $i = 1, \dots, n_{jhg}$, where the number of calls $n_{jhg}$ between home team $j$, visitor $h$, in game $g$ varied from 3 to 27. In total, 4972 calls were reported in the data.

\cite{roback2021beyond} used bootstrapping to obtain a $p$ value for testing a null model where all three random slopes for foul differential were excluded against a full model where all six random effects are included. After running 1000 simulations, a $p$ value of .454 was obtained, from which they concluded to retain their null model which only included the three random intercepts. On our computer (Windows laptop with processor Intel(R) Core(TM) i7 2.80GHz and 16GB RAM), obtaining this $p$ value took approximately 6 hours. Importantly, fitting this model using \textit{\textbf{lme4}} \citep{bates2015package} returned a warning \textit{``Model failed to converge"} for this specific model and data. Therefore, it is difficult to fully trust the results obtained from their example.

\par
To obtained EBFs, we fitted this model using Stan \citep{carpenter2017stan} with flat priors for all parameters. For more details on the Hamiltonian Monte Carlo (HMC) design and the estimates of the random-effect variance parameters see Appendix \ref{app:estimates}. Using the fitted model, we tested all random effects separately against the full model. This only requires fitting the full model. Table \ref{tab:basketball} shows the results of the logarithm of the EBFs for all random effects. From this table, we can see that none of the six random effects received evidence for its inclusion in the full model. As \cite{roback2021beyond} tested a null model with only the random intercepts against the full model (yielding a $p$ value of .454), we also computed the corresponding the log EBF for this comparison which yielded clear evidence for this null model 24.39. To compute this EBF, we combined the estimates of all 340 game level random effects, the 39 home team random effects, and the 39 visitor random random effects, yielding $418$ random effects in total, together with its corresponding joint $418\times 418$ covariance matrix to compute the numerator in \eqref{SDBF}, and a diagonal covariance matrix, diag$(\bar{\tau}^2_1\textbf{1}_{340}',\bar{\tau}^2_2\textbf{1}_{39}',\bar{\tau}^2_3\textbf{1}_{39}')$, to compute the denominator. Note that this log EBF is close to the sum of the EBFs when testing the separate foul differential random effects, but the difference is caused by using the joint covariance matrix of all 418 random effects. Note that the EBF for this joint comparison yields the same conclusion as the $p$ value test reported by \cite{roback2021beyond}. The EBF is more flexible to test many different random effects (and combinations thereof) once the full model has been fitted.

\begin{table}[ht]
\centering
\begin{tabular}{lcc}
  \hline
  \multicolumn{3}{c}{Log of EBFs - Basketball data} \\ 
  \hline
           & foul differential & intercept \\ 
  game     &   13.63   & 86.12 \\ 
  home team &  5.22    & 9.06 \\ 
  visitor &  4.65     & 7.92 \\ 
   \hline
\end{tabular}
\caption{Logarithm of the empirical Bayes factors (EBF) for testing each of the 6 random effects to be fixed against the full random effects model for the basketball data application from \cite{roback2021beyond}.}
\label{tab:basketball}
\end{table}

\subsection{Testing spatial random effects}

\par
Spatial random effects models are frequently employed in spatial statistics and analysis \citep{kang2011bayesian, baltagi2013generalized}. They serve to address spatial autocorrelation, which occurs when spatially close observations exhibit greater similarity than those further apart. These effects prove invaluable in the analysis of spatially correlated datasets, including geographical data and data obtained from spatially distributed entities such as households, census tracts, or grid cells.

\par
\cite{spiegelhalter2002bayesian} explored the data set on cancer lip rates in 56 counties in Scotland. The data contains information on the observed ($y_{j}$) and expected ($x_{j}$) number of lip cancer cases in each county. They assume the observed counts to follow a Poisson distribution, where $e^{\mu_{j}}$ denotes the true area-specific risk of lip cancer, for area $j = 1, \dots, 56$. The model is defined as

\bigskip


\noindent
\begin{equation}
    \begin{gathered}
    \label{eq:spatial_random}
        y_{j} \sim \text{Poisson}\Big( \exp(\mu_{j}(\bm\theta)) x_{j}\Big)\\
        \mu_j(\bm\theta) = \alpha_{0} + \theta_{j1} + \theta_{j2}\\
        \theta_{j1} \sim \mathcal{N}(0, \tau^2_{\theta_1})\\
        \theta_{j2} | \bm{\theta}_{-j2} \sim \mathcal{N}\Bigg(\frac{1}{L_n} \sum_{\ell=1}^{L_n} \theta_{j2}, \frac{\tau^{2}_{\bm{\theta}_2}}{L_n} \Bigg),
    \end{gathered}
\end{equation}

\noindent
where $\alpha_{0}$ is a fixed intercept, $\theta_{n1}$ are exchangeable random effects, $\theta_{n2}$ are spacial random effects and $\bm{\theta}_{-j2}$ is a vector containing all elements of $\bm{\theta}_2$ except the $j$-th component. Moreover, $L_{n}$ is the number of counties adjacent to county $j$, $\tau^{2}_{\bm{\theta}_2}$ is a random-effect variance parameter, and $\bm{\phi} = (\alpha_{0})$. The distribution of $\bm{\theta}_2 = (\theta_{1,2}, \dots, \theta_{56,2})$ implies a structured covariance matrix
.
To see this, we can write the joint distribution of $\bm{\theta}_2$ as

\begin{equation}
    \bm{\theta}_2 = \bm{W} \bm{\theta}_2 + \mathcal{N}(\bm{0}, \tau^{2}_{\bm{\theta}_2} \text{diag}(1/L_{1}, \dots, 1/L_{56})),
\end{equation}

\noindent
where $\bm{W}$ is a weight matrix and $\text{diag}(1/J_{1}, \dots, 1/J_{56})$ is a diagonal matrix with entries corresponding to the number of adjacent counties to each county $j$. After some probability calculus, the joint distribution of $\bm{\theta}_2$ can be written as

\begin{equation}
\label{eq:struct_model}
    \bm{\theta}_2 \sim \mathcal{N}\Big(\bm{0}, \tau^{2}_{\bm{\theta}_2} {(\bm{I} - \bm{W})}^{-1} \bm{B} {({(\bm{I} - \bm{W})}^{-1})}^{T} \Big),
\end{equation}

\noindent
where $\bm{I}$ is an identity matrix with dimensions $56 \times 56$ and $\bm{B}$ is $\text{diag}(1/L_{1}, \dots, 1/L_{56})$. Thus, the model has two random effects where the first has a simple diagonal covariance structure $\tau^2_{\theta_1}\textbf{I}$ and the second has a structured covariance structure as in Equation \eqref{eq:struct_model}.

\par
\cite{spiegelhalter2002bayesian} fitted all four random effects models (by including/excluding the first and second random effect) and computed the respective DICs. The model where both random effects were omitted resulted in the poorest fit while all other three random effects models resulted in an acceptable fit. In the end, the random effects model where the second (spatial) random effect was included and the first random effect was excluded was preferred based the number of effective parameters and the fit of these three models.


\par
To compute the empirical Bayes factors, only the full random effects model in Equation \eqref{eq:spatial_random} needed to be fit. We fitted a fully Bayesian model using Stan with flat priors for all parameters, for information on the HMC design and estimates of the random-effect variances see Appendix \ref{app:estimates}. Table \ref{tab:lipcancer} summarizes the EBFs. The logarithm of the EBF when fixing the first random effect against the full model was equal to $9.55$ implying clear evidence for absence of the first random effect. The logarithm of the EBF for testing the second (spatial) random effect against the full model was equal to $-1.69$ implying some minor evidence towards the inclusion of the spatial random effect. Based on these results, we end up with the same recommendation as \cite{spiegelhalter2002bayesian} where a random effects model is preferred which only includes a spatial random effect.

\begin{table}[ht]
\centering
\begin{tabular}{lcc}
  \hline
  \multicolumn{2}{c}{Log of EBFs - Lip cancer data} \\
  \hline
  random intercept & 9.55 \\ 
  spatial random effect & $-1.69$ \\
   \hline
\end{tabular}
\caption{Logarithm of the Empirical Bayes Factor for the lip cancer data analyzed by \cite{spiegelhalter2002bayesian}.}
\label{tab:lipcancer}
\end{table}

\subsection{Testing individual and temporal measurement invariance in dynamic structural equation models}

\par
Dynamic structural equation models (DSEMs) are statistical models used in the social sciences to analyze time-series data to understand the development of variables over time and their relationships \citep{asparouhov2018dynamic}. Random effects are particularly useful for this class of models to account heterogeneity caused by temporal dependencies and individual dependencies in addition to latent variables. These models extend the traditional structural equation modeling (SEM) framework by explicitly incorporating temporal dynamics into the model.

\cite{mcneish2021measurement} conducted an analysis using a dynamic structural equation model (DSEM) on data of $n = 1, \dots, 50$ overweight/obese adults with binge eating disorder. These subjects were observed during a 28-day period at 4-hour intervals, which amounts to $t = 1, \dots, 152$ intervals. Three items associated with perseverance were used in their analysis. The full DSEM is formulated as follows

\noindent
Model M$_1$:

\begin{equation}
\begin{gathered}
\label{eq:dsem_time_person_rnd_eff}
    \bm{y}_{nt} \sim \mathcal{N}(\bm{\nu}_{nt} + \bm{\Lambda}_{nt} \bm{\eta}_{nt}, \bm{\Sigma})\\
    {\eta}_{nt} = {\eta_w}+{\theta}_{n 1} + {\theta}_{t 4}\\
    \bm{\nu}_{nt} = \bm{\alpha}_{\nu} + \bm{\theta}_{n 2} + \bm{\theta}_{t 5}\\
    \bm{\Lambda}_{nt} = \bm{\alpha}_{\Lambda} + \bm{\theta}_{n 3} + \bm{\theta}_{t 6}\\
    \bm{\theta}_{n} \sim \mathcal{N}(\bm{0}, \bm{\Psi}_{\theta_n})\\
    \bm{\theta}_{t} \sim \mathcal{N}(\bm{0}, \bm{\Psi_{\theta_t}}),\\
\end{gathered}
\end{equation}


\noindent
where {$\eta_w\sim N(0,1)$}, $\bm{y}_{nt}$ is a $3 \times 1$ vector containing the responses of subject $n$ at time $t$. Moreover, $\bm{\nu}_{nt}$ is a $3 \times 1$ vector of intercepts, $\bm{\Lambda}_{nt}$ is a $3 \times 3$ diagonal matrix with factor loadings, $\bm{\eta}_{nt}$ is a $3 \times 1$ vector of latent variables, $\bm{\Sigma}$ is a diagonal matrix, and $\bm{I}$ is an identity matrix. Moreover, $\bm{\alpha}_{\nu}$ and $\bm{\alpha}_{\Lambda}$ are vectors containing intercepts for the general model intercept and the factor loading matrix. Moreover, $\theta_{n 1}$, $\bm{\theta}_{n 2}$, $\bm{\theta}_{n 3}$ are random effects clustering subjects, with $\bm{\theta}_{n 2}$ and $\bm{\theta}_{n 3}$ being $3 \times 1$ vectors. Thus $\bm{\theta}_{n} = (\theta_{n1}, \theta_{n21}, \theta_{n22}, \theta_{n23}, \theta_{n31}, \theta_{n32}, \theta_{n33})'$ and $\bm{\Psi}_{\theta_n} = \text{diag}(\tau^2_{n1}, \tau^2_{n21}, \tau^2_{n22}, \tau^2_{n23}, \tau^2_{n31}, \tau^2_{n32}, \tau^2_{n33})$. Similarly, $\theta_{t 4}$, $\bm{\theta}_{t 5}$, $\bm{\theta}_{t 6}$ are random effects clustering time, with $\bm{\theta}_{t 5}$ and $\bm{\theta}_{t 6}$ being $3 \times 1$ vectors. Hence, $\bm{\theta}_{t} = (\theta_{t1}, \theta_{t21}, \theta_{t22}, \theta_{t23}, \theta_{t31}, \theta_{t32}, \theta_{t33})'$ and $\bm{\Psi}_{\theta_t} = \text{diag}(\tau^2_{t1}, \tau^2_{t21}, \tau^2_{t22}, \tau^2_{t23}, \tau^2_{t31}, \tau^2_{t32}, \tau^2_{t33})$. Therefore, by testing the random effects of the intercepts and loadings across individuals and across measurement occasions (time), we can check individual and temporal measurement invariance.

\par



\par
\cite{mcneish2021measurement} eyeballed the random-effect variance estimates and concluded that the time-level random-effect variance of the factor loadings was very close to zero, and thus, could be excluded, whereas the individual-level random-effect variance was about 0.7 (see Table 2 in \cite{mcneish2021measurement}). Table \ref{tab:bingeeating} shows the logarithm of the EBFs for this illustration. Given that the log EBFs for the time level of the factor loadings (last 3 rows) are all positive, we would reach the same conclusion however using a formal test rather than a potentially subjective eyeball inspection. Therefore, the $\bm{\theta}_{t}$ in the fourth row of Equation \ref{eq:dsem_time_person_rnd_eff} could be excluded to obtain more parsimonious model.

\begin{table}[ht]
\centering
\begin{tabular}{lcc}
  \hline
  \multicolumn{3}{c}{Log of EBFs - Eating disorder data} \\
  \hline
   & person level & time level \\ 
  intercept item 1 & $-384.15$ & $-1117.82$ \\ 
  intercept item 2 & $-502.39$ & $-1433.04$ \\ 
  intercept item 3 & $-277.78$ & $-590.23$ \\ 
  loading item 1 & $-1294.08$ & 28.80 \\ 
  loading item 2 & $-2121.06$ & 30.50 \\ 
  loading item 3 & $-1569.98$ & 31.10 \\  
   \hline
\end{tabular}
\caption{Logarithm of the Empirical Bayes Factor for the binge eating disorder data application from \cite{mcneish2021measurement}.}
\label{tab:bingeeating}
\end{table}


\subsection{Testing random intercepts in cross-lagged panel models}

Cross-lagged panel models are widely used for analyzing reciprocal influences between different psychological constructs over time in longitudinal panel data. In this illustration, we compare the standard Cross-Lagged Panel Model (CLPM) with the Random Intercept Cross-Lagged Panel Model (RI-CLPM), an extension proposed by \cite{hamaker2015critique}.

\par
The RI-CLPM is defined as follows:

\noindent
\begin{eqnarray}
\nonumber    x_{nt} &=& \mu_{t} + \theta_{n1} + p_{nt}\\
\nonumber    y_{nt} &=& \pi_{t} + \theta_{n2} + q_{nt}\\
\nonumber    p_{nt}&=& \alpha_t p_{n, t-1} + \beta_t q_{n, t-1} + u_{nt}\\
\nonumber    q_{nt}&=& \delta_t q_{n, t-1} + \gamma_t p_{n, t-1} + \nu_{nt}\\
\nonumber        \begin{bmatrix}
        \theta_{n1} \\
        \theta_{n2}
    \end{bmatrix} &\sim&
    \mathcal{N}\Bigg(\bm{0}, \bm\Psi(\bm\tau)    
    \Bigg), \text{ with }\bm\Psi(\bm\tau) = \begin{bmatrix}
        \tau^2_{\theta_1} \ \ \tau_{\theta_1\theta_2}\\
        \tau_{\theta_1\theta_2} \ \ \tau^2_{\theta_2}
    \end{bmatrix},
\end{eqnarray}


\noindent
where $x_{nt}$ and $y_{nt}$ represent the two constructs of interest for individual $n$ at measurement wave $t$. The parameters $\mu_{t}$ and $\pi_{t}$ are temporal grand means at wave $t$, while $p_{nt}$ and $q_{nt}$ represent individual temporal deviations from these means. 
The autoregressive parameters $\alpha_t$ and $\delta_t$ capture the stability of each construct over time, and the cross-lagged parameters $\beta_t$ and $\gamma_t$ represent the lagged effects. The residual terms $u_{nt}$ and $\nu_{nt}$ are modelled as $(u_{nt},\nu_{nt}) \sim \mathcal{N}(\bm 0, \bm\Sigma)$, with $\bm\Sigma$ an unstructured covariance matrix. The random intercepts, $\theta_{n1}$ and $\theta_{n2}$, capture stable between-person differences, thereby separating them from within-person processes.

The choice between the CLPM and the RI-CLPM affects parameter estimates and, consequently, the conclusions drawn from the data. If stable individual differences are present, applying the CLPM may lead to biased parameter estimates \citep{hamaker2015critique}. Conversely, if such differences are negligible, applying the RI-CLPM may overfit the data. According to the principle of parsimony (Occam’s razor), the simpler model (i.e., CLPM) should be preferred when its fit is comparable to that of a more complex model (i.e., RI-CLPM). This is also related to the bias-variance trade-off: including unnecessary parameters, such as random intercepts, can increase the variance and reduce the predictive power.

To illustrate the application of the EBF for testing the necessity of random intercepts, we used data from \cite{mackinnon2022tutorial}, which included responses from 251 participants collected over five days. The two constructs measured were perfectionistic self-presentation (PSP), assessed using three items, and state social anxiety (SSA), assessed using seven items. Our analysis was based on (averaged) observed scores of the indicators, following the model proposed by Hamaker et al. (2011). Importantly, the proposed EBF approach can be straightforwardly extended to more complex RI-CLPMs with latent variables (as in Mackinnon et al. 2022), as well as to other extensions \citep{mulder2021three}. 

We fitted the RI-CLPM using the R package `blavaan' \citep{merkle2015blavaan, merkle2020efficient}. To account for the dependency between the two random intercepts, we computed the EBF to assess the joint inclusion of both random intercepts, i.e., whether $\theta_{n1}$ and $\theta_{n2}$ are both equal to zero. Note that the stacked vector of the two random intercepts, $(\bm{\theta}_{1}',\bm{\theta}_{2}')'$, has a structured covariance matrix that is equal to $ \bm\Psi(\bm\tau) \otimes \textbf{I}_J$. The logarithm of the EBF was $-761.59$, implying strong evidence for the inclusion of both random intercepts. Furthermore, Table \ref{tab:riclpm_EBF} provides the EBFs for testing the exclusion of the random effects against the full model separately. Both indicate clear evidence in favor of including the random intercepts.

\begin{table}[ht]
\centering
\begin{tabular}{lcc}
  \hline
  \multicolumn{2}{c}{Log of EBFs - Random intercepts cross-lagged panel model} \\
  \hline
  first random intercept & -738.42 \\ 
  first random intercept & -267.34 \\ 
   \hline
\end{tabular}
\caption{Logarithm of the Empirical Bayes Factor for the tree growth data \cite{baey2023vartestnlme}.}
\label{tab:riclpm_EBF}
\end{table}

We then compared the conclusion of the proposed EBF with the conclusion when using chi-bar-square test \citep{kuiper2020,hamaker2015critique, stoel2006likelihood}, a significance test based on a weighted sum of chi-square distributions. When the $p$-value falls below a pre-specified significance level the null model, i.e., the CLPM, is rejected in favor of the more complex RI-CLPM. For these data, a $p$-value of approximately 0 was obtained implying a significant result for any conventional significance level. Thus, the significance test also implies that the CI-CLMP is preferred.

It is important to note however that this significance test can result in inflated type I errors under a true CLPM implying that a correct CLPM will be rejected more often than the chosen significance level thereby jeopardizing the reliability of the test. As demonstrated in a simulation (Appendix \ref{app:chi_squared}), the distribution of $p$-values under the null deviates from a uniform distribution but, in fact, is more concentrated around 0. This indicates that the chi-bar-square test fails to properly control the Type I error rate, rejecting a correctly specified CLPM too often on average. This is especially problematic as significance tests are already known to be too liberal \citep{berger1987testing,sellke2001calibration,wagenmakers2007practical}.

\subsection{Testing random effects in nonlinear mixed effects models}
Nonlinear mechanisms between variables are ubiquitous in scientific practice. When the data are clustered and when a parameterized function of the nonlinear mechanism is known, e.g., based on certain natural laws, nonlinear mixed effects models are useful \citep[][Ch. 19]{pinheiro2007linear,gelman2014bayesian}. When building such nonlinear mixed effects model, a key question is again whether the degree of heterogeneity of each parameter across clusters is sufficiently large in order to include it as a random effect.

To illustrate how the proposed EBF can be used for this purpose, we consider an application that presented by \citep{baey2023vartestnlme} who used a $p$ value for assessing for the presence of random effects using bootstrapping. A nonlinear mixed effects model was considered for studying the growth of Loblolly pine trees \citep{kung1986fitting} where the growth of 14 trees was recorded on 6 occasions between the ages of 3 and 25 years. The following nonlinear model with three independent random effects was considered:
\begin{eqnarray*}
y_{ij} &=& \eta_{1i} + (\eta_{2i}-\eta_{1i})\exp\{-e^{\eta_{3i}x_{ij}}\}+\epsilon_{ij}\\
(\eta_{1i},\eta_{2i},\eta_{3i})' &=& (\mu_{1},\mu_{2},\mu_{3})' + (\theta_{1i},\theta_{2i},\theta_{3i})'\\
(\theta_{1i},\theta_{2i},\theta_{3i})'&\sim& \mathcal{N}((0,0,0)',\text{diag}(\tau_1^2,\tau_2^2,\tau_3^2))\\
\epsilon_{ij}&\sim& \mathcal{N}(0,\sigma^2)
\end{eqnarray*} 
where $y_{ij}$ is the length of the $i$-th tree at measurement $j$ and $x_{ij}$ is its corresponding age. The three random effects were labeled `Asym', `R0', `lrc', respectively. The full model was fitted using brms \citep{burkner2017brms}. Flat priors were used for the fixed effects and random effects variances. The logarithm  of EBFs in favor of absence against presence of each random effect against the full model were equal to 4.37, 3.27, and 4.12, respectively, which imply evidence in favor of absence of the random effects. The bounds of the $p$ value as reported by \citep{baey2023vartestnlme} for testing a null model which only included the first random effect against the full model were equal to .05 and .20, implying a nonsignificant result using a significance threshold of .05. The EBF for this null model against the full model yielded 7.19 (note that it is not exactly equal to $3.27 + 4.12$ because now the joint posterior covariance matrix of the two random effects. Thus, the EBF yields clear evidence in favor of this null model against the null model, a conclusion which is similar to the conclusion based on the $p$ value of \citep{baey2023vartestnlme}. 

\begin{table}[ht]
\centering
\begin{tabular}{lcc}
  \hline
  \multicolumn{2}{c}{Log of EBFs - Growth data} \\
  \hline
  random `Asym' effect & 4.37 \\ 
  random `R0' effect & 3.27 \\ 
  random `lrc' effect & 4.12 \\
   \hline
\end{tabular}
\caption{Logarithm of the Empirical Bayes Factor for the tree growth data \cite{baey2023vartestnlme}.}
\end{table}

\section{Discussion and conclusion}\label{sec:discussion}


\par
In empirical research, the collected data are often clustered, e.g., due to hierarchical (multilevel) structures \citep[e.g.,][]{roback2021beyond,spiegelhalter2002bayesian,mcneish2021measurement}. When building statistical models for such data, an important question is whether the implied heterogeneity that is caused by the clustered structure is sufficiently compelling for random effects to be included in the model. Because of the ubiquity of this problem in statistical practice, many approaches have been proposed in the statistical literature to address this question including eyeballing statistical descriptive results, significance tests, cross-validation, information criteria or (regular) Bayes factors \citep{crainiceanu2008likelihood,zhang2008variance,hamaker2011model,mulder2013bayesian,berrar2019cross}. As discussed in this paper, these methods have different advantages and disadvantages. 

\par
The current paper proposed a novel empirical Bayes factor (EBF) for this testing problem which avoids certain limitations of existing methods. Its two main advantages are its simplicity to compute it (only requiring estimates of the random effects, the uncertainty, and the lower level variance estimates) and its flexibility to test structural heterogeneity under any type of mixed effects model. Moreover, the methodology avoids the need for manual prior specification \cite[thereby abiding notions of empirical Bayes,][]{casella1985introduction} by using estimated variance components on the second level to serve as prior variances for the random effects on the first level. Computationally expensive methods, such as marginal likelihoods, bootstrap simulations, or cross-validation, are avoided by adopting a Savage-Dickey density ratio together with Gaussian approximations of the posterior. Consequently, only the full mixed effects model needs to be fitted in order to compute the EBFs for all separate random effects which is not the case for existing alternatives. 

\par
A synthetic data study assessed the behavior of the EBFs in controlled setting when using the posterior mean as plug-in estimate of the lower level variance components or when using the full posterior. The degree of heterogeneity of one random effect was varied in addition to the number of clusters and the cluster sizes. When using the posterior mean to compute EBFs, the anticipated selection behavior was obtained. When using the full posterior to compute the EBF, the EBF was considerably biased towards the full model. For this reason, computing EBFs using the posterior mean is recommended.

\par
Five different empirical applications under different mixed effects models were used to illustrate the flexibility of the approach, namely, generalized linear mixed models (GLMMs) with cross random effects, spatial random-effects models, dynamic structural equation models, random intercepts in cross-lagged panel models, and nonlinear mixed effects models. These applications showed that test often results in comparable selection behavior as existing methods. Of course, there are also instances where the outcome would be different. For example, Bayes factors are known to be relatively conservative towards a null model in comparison to other approaches \citep[e.g.,][]{berger1987testing,weakliem1999critique,mulder2024bayesian}. Despite the (dis)similarities with other approaches, the proposed EBFs has the advantage of being easy to apply and computationally fast when testing random effects in mixed effects models.

\par
The EBF relies on the Savage-Dickey density ratio which has received some attention in the literature \citep{verdinelli1995computing,wetzels2010encompassing,marin2010resolving,heck2019caveat,mulder2022generalization}. Importantly, in order for the Savage-Dickey density ratio to be equal to a Bayes factor, the prior for the nuisance parameters under the restricted model must be equal to the prior under the full model by conditioning on the null restriction that the random effects are zero. When the random effects are modelled independently, this condition automatically holds. When the condition does not hold one could consider alternative priors for the nuisance parameters. We leave this topic for future work.

\par
Another potential direction for future work would be to use the EBF to search for the best fitting random effects model in an iterative manner. After fitting the full mixed effects model, one could omit the random effect which received most support to be excluded, and refit the model. Subsequently, EBFs can be computed for testing the remaining random effects against this reduced full model. This iterative process can be repeated until a full model is obtained for which all EBFs suggest that the remaining random effects should be included. We leave the study of this iterative search for future research.

\par
Finally we note that EBFs can also be computed using classical output, e.g., from `lme4' \citep{bates2015package}, or using Bayesian posterior modes using `blme' \citep{dorie2015package}. Studying the implied behavior of the EBF when using output from these packages are left for future work.


\appendix


\section{Additional plots of the simulation}\label{AppFig}
\begin{figure}[thp]
    \centering
    \includegraphics[width = 10cm]{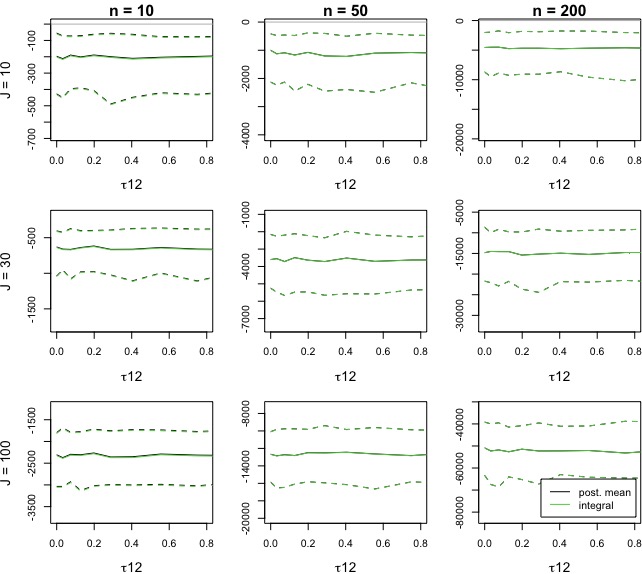}\\
    \caption{The 5\%, 50\%, and 95\% quantiles of the sampling distribution of the logarithm of the EBF in favor of absence against presence of the second random effect of the first crossed dimension (i.e., $\tau_{12}=0$ versus $\tau_{12}>0$) as a function of the true data generating $\tau_{11}$ for different cluster sizes $n$ and different numbers of clusters $J$ when using the posterior mean (black lines) or the full posterior (green lines) in \eqref{EBF01final}.}
    \label{fig_sim_tau12}
\end{figure}

\begin{figure}[thp]
    \centering
    \includegraphics[width = 10cm]{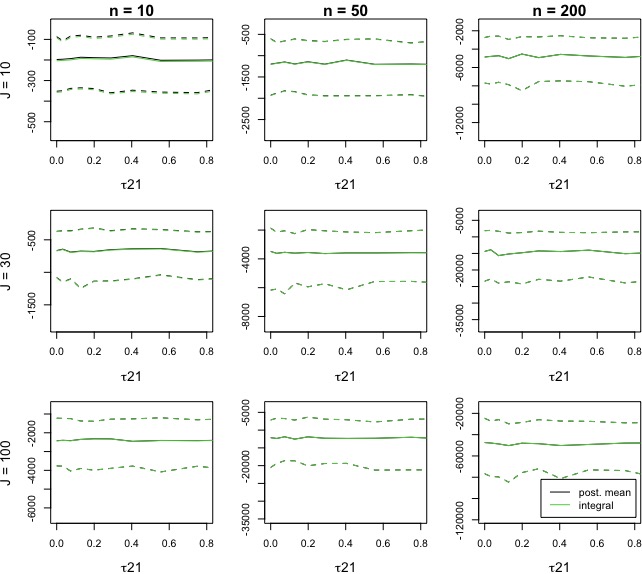}\\
    \caption{The 5\%, 50\%, and 95\% quantiles of the sampling distribution of the logarithm of the EBF in favor of absence against presence of the first random effect of the second crossed dimension (i.e., $\tau_{21}=0$ versus $\tau_{21}>0$) as a function of the true data generating $\tau_{11}$ for different cluster sizes $n$ and different numbers of clusters $J$ when using the posterior mean (black lines) or the full posterior (green lines) in \eqref{EBF01final}.}
    \label{fig_sim_tau21}
\end{figure}

\begin{figure}[thp]
    \centering
    \includegraphics[width = 10cm]{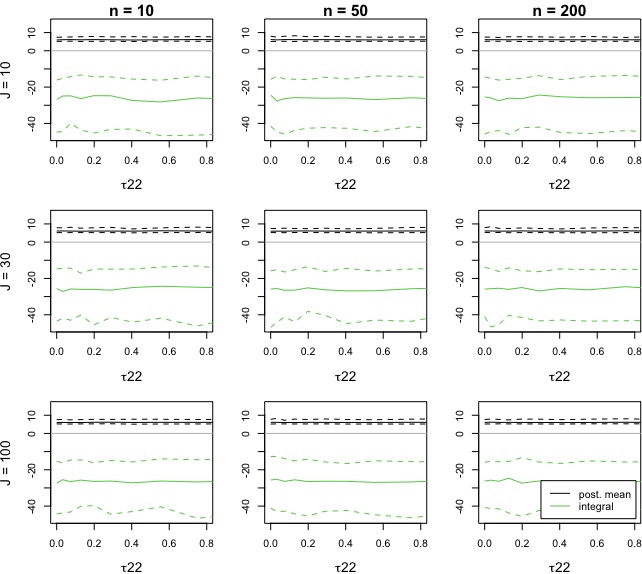}\\
    \caption{The 5\%, 50\%, and 95\% quantiles of the sampling distribution of the logarithm of the EBF in favor of absence against presence of the second random effect of the second crossed dimension (i.e., $\tau_{22}=0$ versus $\tau_{22}>0$) as a function of the true data generating $\tau_{11}$ for different cluster sizes $n$ and different numbers of clusters $J$ when using the posterior mean (black lines) or the full posterior (green lines) in \eqref{EBF01final}.}
    \label{fig_sim_tau22}
\end{figure}

\begin{figure}[thp]
    \centering
    \includegraphics[width = 13.5cm]{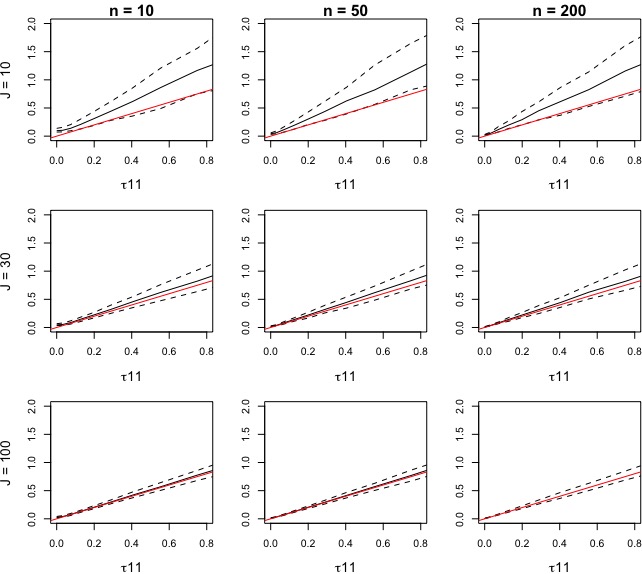}\\
    \caption{The 5\%, 50\%, and 95\% quantiles of the sampling distribution (solid and dotted black lines) of the posterior mean of $\tau_{11}$ as a function of the true data generating $\tau_{11}$ (red line) for different cluster sizes $n$ and different numbers of clusters $J$.}
    \label{fig_sim_tau_postmean}
\end{figure}

This appendix contains the plots of the quantiles of the sampling distribution of the EBF for testing the second random effect of the first crossed dimension ($\tau_{12}$) (Figure \ref{fig_sim_tau12}), for testing the first random effect of the second crossed dimension ($\tau_{21}$) (Figure \ref{fig_sim_tau21}), and for testing the second random effect of the second crossed dimension ($\tau_{22}$) (Figure \ref{fig_sim_tau22}) as function of the data generating $\tau_{11}$ from the simulation study (Section \ref{sec:simulation}). Figure \ref{fig_sim_tau_postmean} shows the quantiles of the sampling distribution of the posterior mean of $\tau_{11}$.

\section{Additional results for the empirical applications}
\label{app:estimates}

In this appendix, we provide the posterior mean estimates for the variance components of the random-effect models fitted in Section \ref{sec:application}. For each application, we will discuss the MCMC setup used to fit those models and provide the estimates for the parameters of interest (i.e. the random-effect variances). The R code for the analyses are provided in the supplementary material.

\subsection{Testing cross random effects in GLMMS}

This is the model fitted in Subsection 5.1, where the example using the basketball data from \cite{roback2021beyond} was discussed. In this application, we ran 5000 iterations of a Hamiltonian Monte Carlo (HMC). The advantage of using HMC is that it requires considerably less iterations to reach convergence and it is not plagued by the random walk behavior that causes severe autocorrelation in posterior samples like traditional MCMC algorithms such as the Gibbs sampler \citep{hoffman2014no}. We discarded the first 2500 samples and kept the other 2500. 

\begin{figure}
    \centering
    \includegraphics[width=1\linewidth]{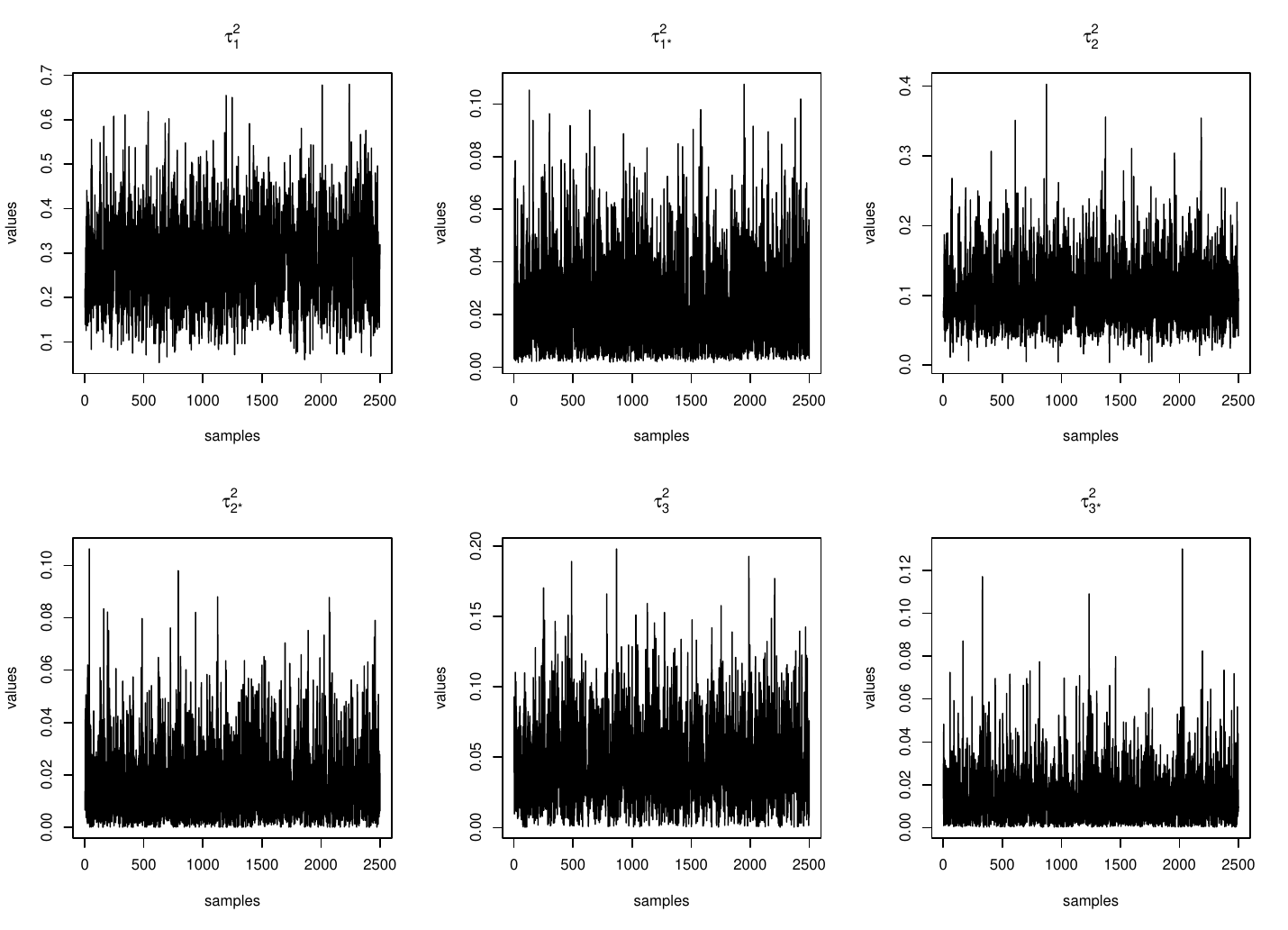}
    \caption{Trace plots of the variance parameters obtained with the Baskteball data.}
    \label{fig:basketball}
\end{figure}

Table \ref{tab:basketball2} shows the posterior point estimates and 95\% posterior credible intervals. Also, figure \ref{fig:basketball} displays trace plots of the posterior samples for the random-effect variance components, the chains have reached convergence.

\begin{table}[ht]
\centering
\begin{tabular}{lcc}
  \hline
  parameter & posterior mean (95\% CI)\\ 
  \hline
    $\tau^2_1$ & 0.2782 (0.1119, 0.4900) \\ 
    $\tau^2_2$ & 0.1024 (0.0339, 0.2165) \\ 
    $\tau^2_3$ & 0.0458 (0.0022, 0.1175) \\ 
    $\tau^2_1*$ & 0.0227 (0.0030, 0.0680) \\ 
    $\tau^2_2*$ & 0.0159 (0.0002, 0.0538) \\ 
    $\tau^2_3*$ & 0.0142 (0.0006, 0.0486) \\ 
   \hline
\end{tabular}
\caption{Random-effect variance posterior estimates with 95\% intervals for the Basketball data.}
\label{tab:basketball2}
\end{table}

\subsection{Testing spatial random effects}

\begin{figure}
    \centering
    \includegraphics[width=1\linewidth]{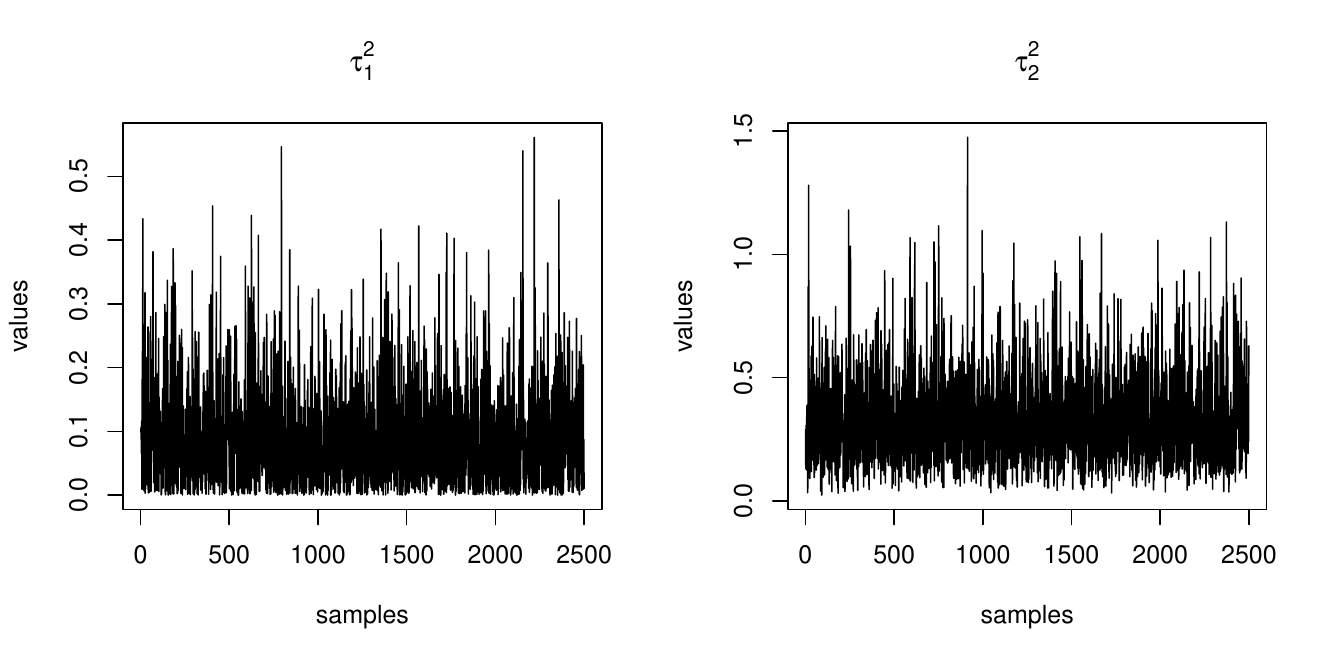}
    \caption{Trace plots of the variance parameters obtained with the Scottish lip cancer data.}
    \label{fig:scotlip}
\end{figure}

This model was fitted in Subsection 5.2 and features the data on the cancer lip rates in Scotland \citep{spiegelhalter2002bayesian}. One again, we ran 5000 iterations of a Hamiltonian Monte Carlo, discarding the first 2500 and saving the last 2500 samples. Table \ref{tab:scotlip} shows the posterior mean and 95\% posterior intervals for the random-effect variance components and Figure \ref{fig:scotlip} displays trace plots of the posterior samples for these parameters.

\begin{table}[ht]
\centering
\begin{tabular}{lc}
  \hline
    parameter & posterior mean (95\% CI) \\ 
  \hline
    $\tau^2_{1}$ & 0.0878 (0.0012, 0.2799) \\ 
    $\tau^2_{2}$ & 0.3300 (0.0719, 0.7598) \\ 
   \hline
\end{tabular}
\caption{Random-effect variance posterior estimates with 95\% intervals for the Scottish lip cancer data.}
\label{tab:scotlip}
\end{table}

\subsection{Testing individual and temporal measurement invariance in dynamic structural equation models}

\begin{figure}
    \centering
    \includegraphics[width=1\linewidth]{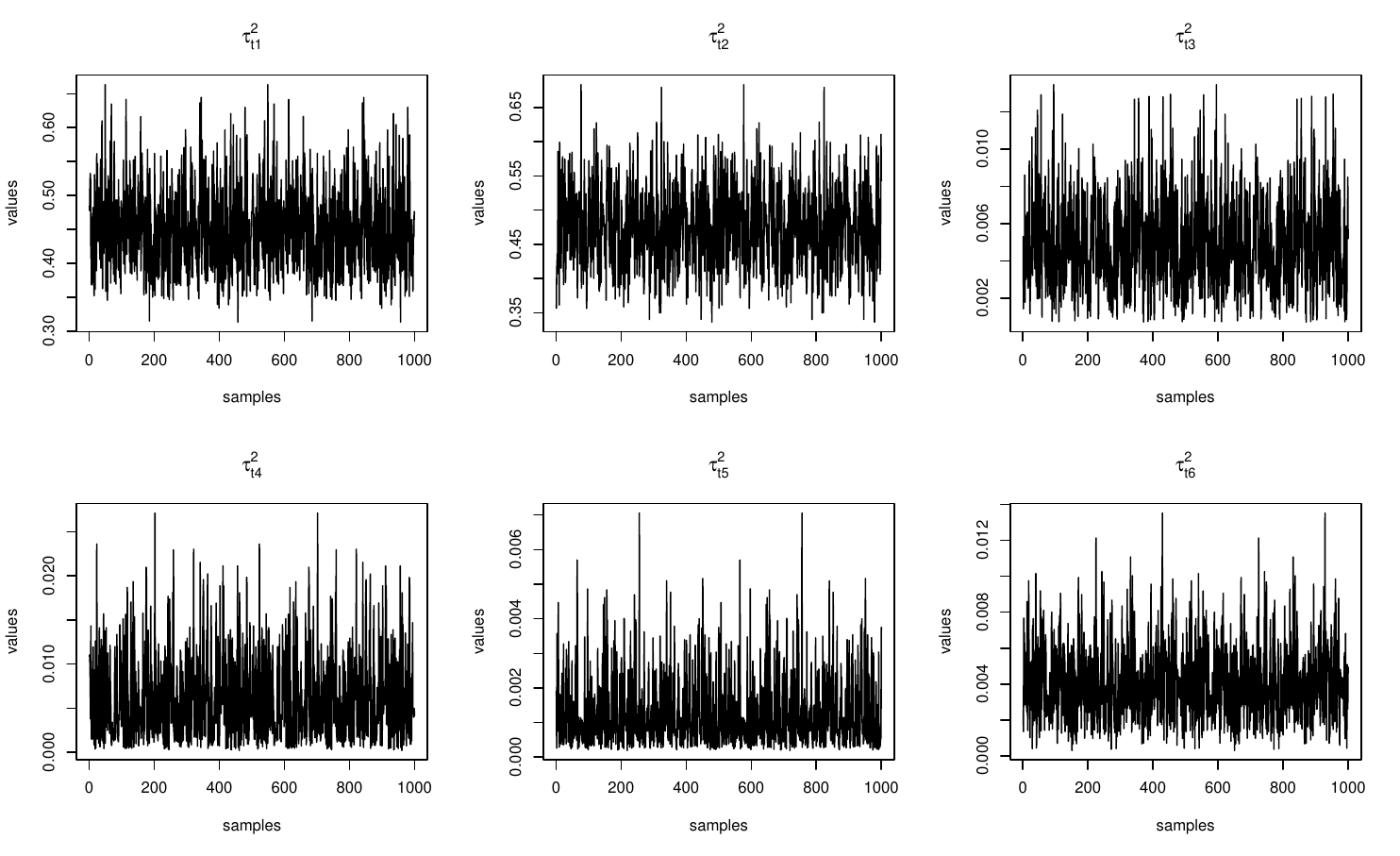}
    \caption{Trace plots of the variance parameters obtained with the Binge eating disorder data.}
    \label{fig:binge_1}
\end{figure}

\begin{figure}
    \centering
    \includegraphics[width=1\linewidth]{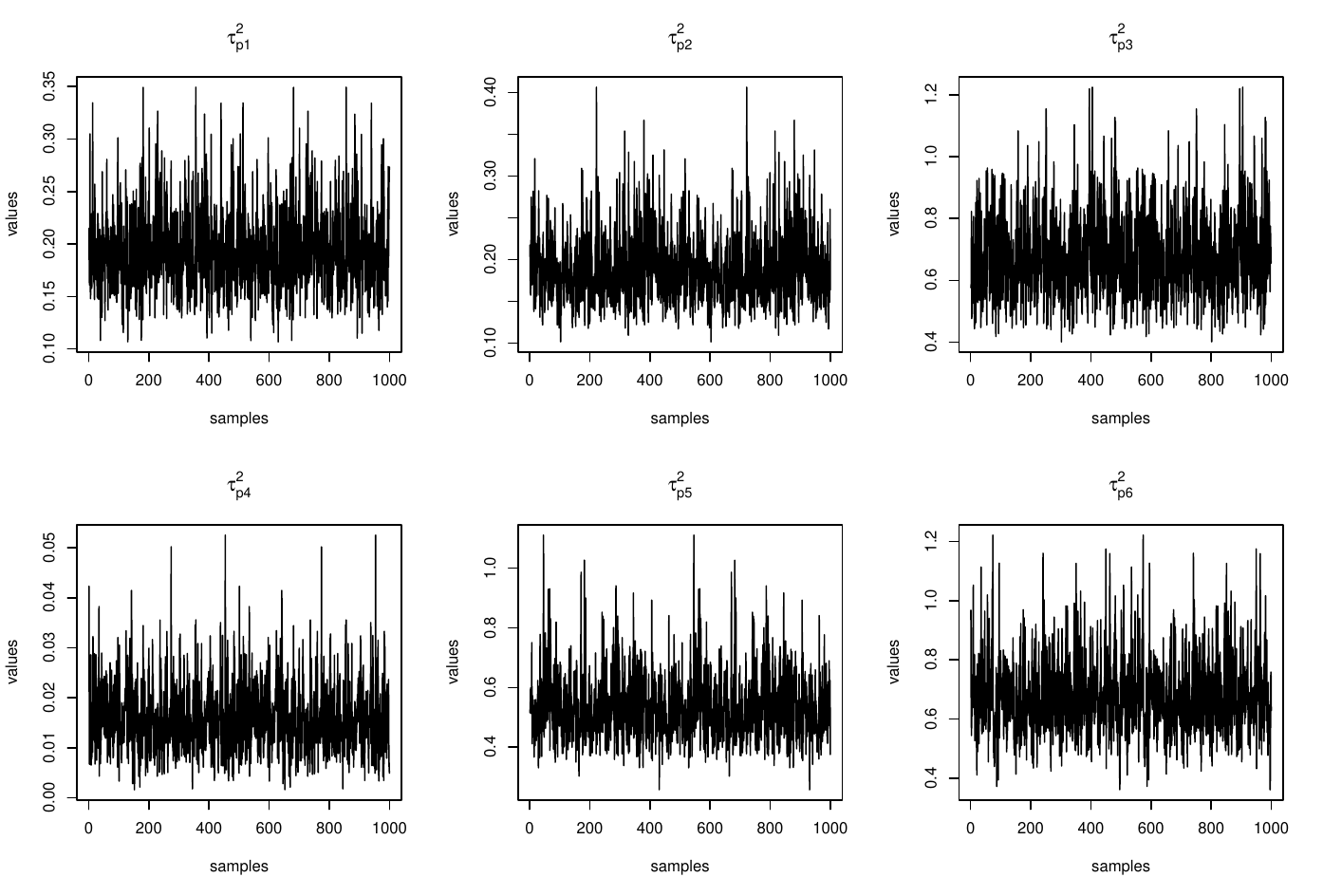}
    \caption{Trace plots of the variance parameters obtained with the Binge eating disorder data.}
    \label{fig:binge_2}
\end{figure}

The DSEM was used in Subsection 5.3, in the application with binge eating disorder \citep{mcneish2021measurement}. For this model, we ran 2000 iterations of an HMC, discarding the first 1000 and keeping the last 1000 samples. Figures \ref{fig:binge_1} and \ref{fig:binge_2} show the trace plots of the posterior samples for the parameters of interest on time and person level, respectively. The chains have converged. Table \ref{tab:binge1} shows posterior mean estimates and 95\% posterior intervals for all random-effect variance parameters.

\begin{table}[ht]
\centering
\begin{tabular}{lc|lc}
  \hline
  \multicolumn{2}{c}{Time Level} & \multicolumn{2}{c}{Person Level}\\
  \hline
  parameter (time) & posterior mean (95\% CI) & parameter (person) & posterior mean (95\% CI) \\ 
  \hline
  $\tau^2_{t1}$ & 0.2950 (0.2194, 0.3729) & $\tau^2_{p1}$ & 0.5312 (0.3616, 0.8352) \\ 
  $\tau^2_{t2}$ & 0.4514 (0.3504, 0.5892) & $\tau^2_{p2}$ & 0.6773 (0.4517, 0.9996) \\ 
  $\tau^2_{t3}$ & 0.4759 (0.3694, 0.5984) & $\tau^2_{p3}$ & 0.6749 (0.4468, 1.0082) \\ 
  $\tau^2_{t4}$ & 0.0013 (0.0002, 0.0040) & $\tau^2_{p4}$ & 0.1889 (0.1220, 0.3002) \\ 
  $\tau^2_{t5}$ & 0.0040 (0.0008, 0.0092) & $\tau^2_{p5}$ & 0.1951 (0.1292, 0.2918) \\ 
  $\tau^2_{t6}$ & 0.0049 (0.0010, 0.0108) & $\tau^2_{p6}$ & 0.1424 (0.0889, 0.2148)\\ 
   \hline
\end{tabular}
\caption{Random-effect variance posterior estimates with 95\% intervals for the Binge eating disorder data.}
\label{tab:binge1}
\end{table}

\subsection{Testing random intercepts in random intercepts cross lagged panel model}

\begin{figure}
    \centering
    \includegraphics[width=0.8\linewidth]{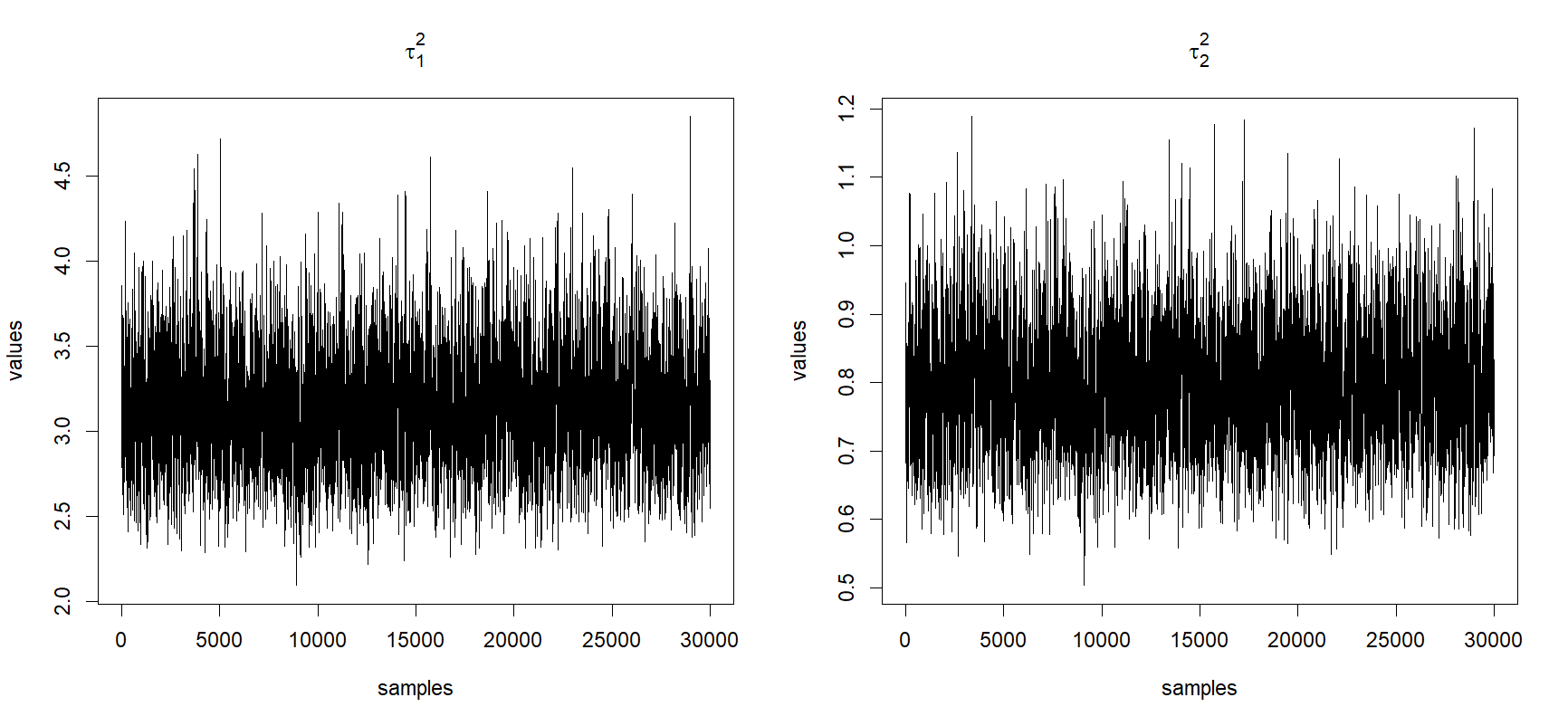}
    \caption{Trace plots of the variance parameters obtained with the data from Mackinnon et al. (2022).}
    \label{fig:ri-clpm}
\end{figure}

The RI-CLPM was fitted using Gibbs sample from the R package blavaan. Table \ref{tab:ri-clpm} shows the random effect variance components estimates and their respective 95\% posterior intervals. Figure \ref{fig:ri-clpm} shows the trace plots of the random effects variances.

\begin{table}
\renewcommand{\thetable}{4}
\centering
\begin{tabular}{lcc}
\hline
parameter & posterior mean (95\% PI)\\
\hline
$\tau^2_1$ & 3.138 (2.573, 3.810) \\
$\tau^2_2$ & 0.795 (0.642, 0.972) \\
\hline
\end{tabular}
\caption{Random-effect variance posterior estimates with 95\% intervals for the data
application from Mackinnon et al. (2022).}
\label{tab:ri-clpm}
\end{table}

\subsection{Testing random effects in a nonlinear mixed effects model}

\begin{figure}
    \centering
    \includegraphics[width=12cm]{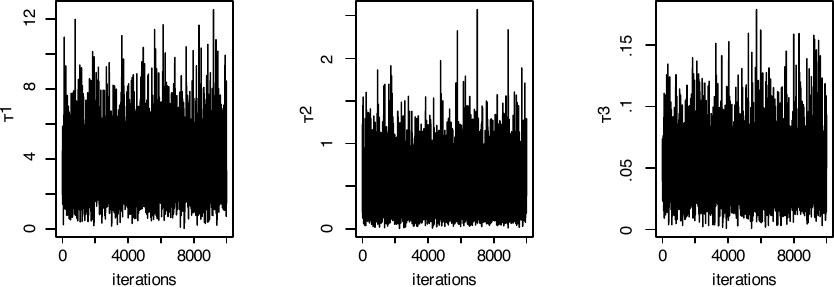}
    \caption{Trace plots of the random effects standard deviations of the nonlinear mixed effects model.}
    \label{fig:tracenonlinear}
\end{figure}

The nonlinear mixed effects model as described in Subsection 5.5 was fitted using brms. Figures \ref{fig:tracenonlinear} shows the trace plots of the posterior samples for the three random effects standard deviations. Table \ref{tab:binge2} show posterior mean estimates and 95\% posterior intervals for all random effects standard deviations.

\begin{table}[ht]
\centering
\begin{tabular}{lcc}
\hline
&posterior mean & 95\% CI\\
  \hline
  $\tau_{1}$ & 3.727 &(1.140, 7.089)  \\ 
  $\tau_{2}$ & 0.485 &(0.090, 1.108)  \\ 
  $\tau_{3}$ & 0.049 & (0.014, 0.096)  \\ 
   \hline
\end{tabular}
\caption{Posterior means and 95\% credible intervals for the three random effects standard deviations for the tree growth data.}
\label{tab:binge2}
\end{table}

\section{Simulation study: Evaluating the sampling distribution of p-values of the chi-bar-square test}\label{app:chi_squared}

To evaluate the performance of the $\bar{\chi}^2$ test (Kuiper, 2020) to control type I error rates when testing the CLPM against the RI-CLPM, we conducted a simulation study. We examined its performance using different sample sizes and different numbers of waves. Data generation and analyses were conducted in R (R Core Team, 2021), using the lavaan (Rosseel, 2012) and blavaan (Merkle \& Rosseel 2015; Merkle et al. 2020) packages, along with the `ChiBarSq.DiffTest' (Kuiper, 2020). 

To generate bivariate longitudinal data with different numbers of waves, we used the `simulateData()' function from `lavaan'. Data were generated under a CLPM model. We varied two factors in the simulation: (1) sample size: 50, 100, 250, 500, and 1000; and (2) number of waves: 3, 5, and 10. This resulted in 15 conditions. For each condition, we generated 200 data sets, leading to a total of 3,000 data sets. Due to convergence issues, the final simulation study included 2,945 data sets.

Figure \ref{fig_sim_chibar} below shows the distributions of $p$-values across all conditions when the data were generated under the null model. Across all combinations of sample sizes and numbers of waves, none of these $p$-values followed a uniform distribution (displayed as red dashed lines) but concentrated around 0. This indicates that the test is unable to properly control the type I error as it will be prone to falsely rejecting the null model too often.

\begin{figure}[ht]
\centering
    \includegraphics[width = 5.5in, height = 2.5in]{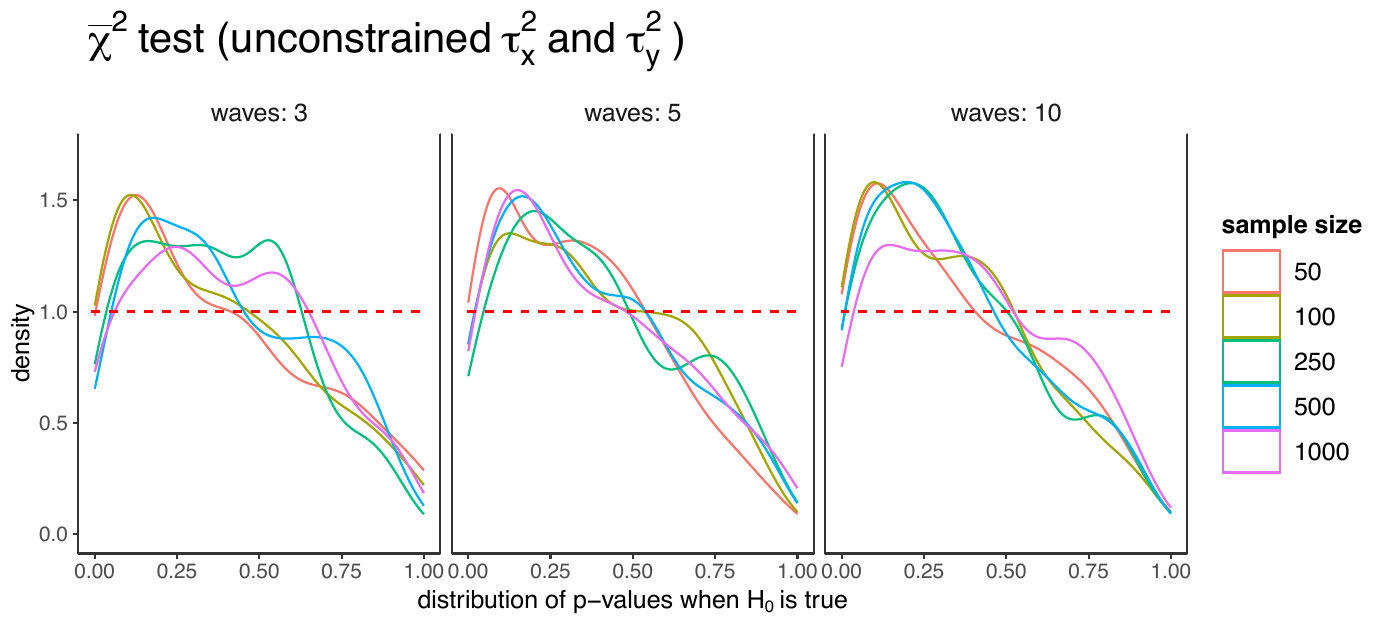}
    \caption{The p-values distribution from the chi-bar-square test when $\tau_x^2=\tau_y^2=0$. They are classified according to the number of waves and sample size. The dashed lines represent the mean of p-values. The data were analyzed in lavaan without constraints on  $\tau^2$ (where the estimates can be negative). None of the plots show a uniform distribution.}
    \label{fig_sim_chibar}
\end{figure}


\clearpage 
\bibliographystyle{apacite}
\bibliography{references}



\end{document}